\documentclass[journal]{IEEEtran}
\ifCLASSINFOpdf
\else
\fi
\usepackage{amsmath} 
\usepackage{amssymb}
\usepackage[ruled,vlined]{algorithm2e}
\usepackage{enumitem}
\usepackage{cite}
\usepackage{graphicx}
\usepackage{subfig}
\usepackage{multirow}

\begin{document}
\title{A Secure Federated Learning Framework for Residential Short Term Load Forecasting}

\author{Muhammad Akbar Husnoo,
        Adnan Anwar,~\IEEEmembership{Member},
        Nasser Hosseinzadeh,~\IEEEmembership{Senior Member},
        Shama Naz Islam,~\IEEEmembership{Member},
        Abdun Naser Mahmood,~\IEEEmembership{Senior Member} and
        Robin Doss,~\IEEEmembership{Senior Member}.% <-this % stops a space
\thanks{This document is the results of the research project funded by the Centre for Cyber Security Research and Innovation (CSRI), School of Information Technology, Deakin University.}% <-this % stops a space
\thanks{M. A. Husnoo, A. Anwar and R. Doss are with the Centre for Cyber Security Research and Innovation (CSRI) at Deakin University.}% <-this % stops a space
\thanks{S. N. Islam and N. Hosseinzadeh are with the Centre for Smart Power and Energy Research (CSPER) at Deakin University.}
\thanks{A. N. Mahmood is with the Department of Computer Science \& IT at Latrobe University.}
}


\markboth{Journal of \LaTeX\ Class Files,~Vol.~14, No.~8, August~2015}%
{Shell \MakeLowercase{\textit{et al.}}: Bare Demo of IEEEtran.cls for IEEE Journals}



\maketitle

\begin{abstract}
Smart meter measurements, though critical for accurate demand forecasting, face several drawbacks including consumers' privacy, data breach issues, to name a few. Recent literature has explored \textit{Federated Learning} (FL) as a promising privacy-preserving machine learning alternative which enables collaborative learning of a model without exposing private raw data for short term load forecasting. Despite its virtue, standard FL is still vulnerable to an intractable cyber threat known as \textit{Byzantine attack} carried out by faulty and/or malicious clients. Therefore, to improve the robustness of federated short-term load forecasting against Byzantine threats, we develop a state-of-the-art differentially private secured FL-based framework that ensures the privacy of the individual smart meter's data while protect the security of FL models and architecture. Our proposed framework leverages the idea of gradient quantization through the Sign Stochastic Gradient Descent (SignSGD) algorithm, where the clients only transmit the `sign' of the gradient to the control centre after local model training.  As we highlight through our experiments involving benchmark neural networks with a set of Byzantine attack models, our proposed approach mitigates such threats quite effectively and thus outperforms conventional FedSGD models.
\end{abstract}

% Note that keywords are not normally used for peer review papers.

% \fi
\IEEEpeerreviewmaketitle
%\section{Preliminary}
%\IEEEPARstart{T}{his} demo file is intended to serve as a ``starter file''
%for IEEE journal papers produced under \LaTeX\ using

\section{Introduction}

\IEEEPARstart{R}{ecent} advances within the Smart Grid (SG) paradigm are geared towards the incorporation of several Internet of Things (IoT) based devices and advanced computing technologies to ensure reliability, flexibility and efficiency of critical power systems \cite{Lamnatou_Chemisana_Cristofari_2022}. With the prevalence of Artificial Intelligence (AI), the enormous amount of highly granular power-related data generated by such intelligent devices enable energy service providers to improve load forecasts, maximize financial gains, devise effective demand management and other grid operation strategies, etc \cite{Sakhnini_Karimipour_Dehghantanha_Parizi_Srivastava_2021}. Besides, consumers can experience better quality of service through personalization of the power system applications and tools \cite{9478223}. In recent years, several decentralized load forecasting solutions are being actively proposed by researchers in the SG domain. Such approaches rely on the sharing of data among several decentralized nodes during the training process to improve accuracy and robustness. However, data sharing raises privacy concerns, even though it can significantly enhance performance. In such cases, the sharing of fine-grained load consumption profiles collected from individual smart meters to central data servers imposes several privacy concerns to energy data owners \cite{husnoo2021false, reda2021taxonomy}. For instance, several studies \cite{10.1145/1878431.1878446, 10.1145/2528282.2528295} have highlighted that simple analysis of load consumption patterns recorded by smart meters can reveal household occupancy rates, the presence of people within a house, and sleep/wake-up time of residents, without any prior knowledge. Indeed, higher resolution of smart meter data leads to higher granularity in information and allows third parties to infer more sensitive information about households.

In such a scenario, Federated Learning (FL) emerges as a viable privacy-preserving distributed computing alternative which transfers computation to energy data owners and allows the training of a global model through collaboration of devices without requiring the migration of data to a central repository for model training \cite{9084352}. Typically, edge devices in an energy system network iteratively train a local model and update the resulting parameters to a central aggregator which accumulates and processes the parameters and then sends back the updated parameters to the edge devices. The communication rounds continue until successful convergence of the model. In spite of the privacy preservation benefits due to the omission of raw data sharing requirements, FL is also efficient in terms of communication resource usage and has higher scalability \cite{9084352}. Recently, FL has gained much attention from researchers to explore its potential benefits within several smart grid domains, namely short-term load forecasting \cite{husnoo2022fedrep, 9148937}, energy theft detection \cite{9531953}, to name a few. Nevertheless, despite its promising privacy-preserving potentials, recent literature has revealed that FL may fail to provide sufficient privacy guarantees in certain circumstances. For example, researchers have discovered that they are able to reconstruct the original raw data from the sharing of gradients of the model during iterations \cite{iDLG}. Furthermore, due to the distributed nature of FL, it is vulnerable to Byzantine faults/attacks whereby the client nodes behave arbitrarily which may be a result of adversarial manipulations or software/hardware faults \cite{FLTrust}. Therefore, it is imperative to design FL mechanisms that are fault-tolerant to such behaviours, provide good generalisation performance and are communication efficient. Consequently, we investigate this research gap in the field of smart grids by contributing to the following:

\begin{enumerate}[leftmargin=*]
    \item Inspired by the idea of gradient quantization, we develop a state-of-the-art privacy-preserving FL-based framework that leverages the SIGNSGD algorithm to improve the robustness of FL strategies for residential short-term load forecasting against Byzantine attacks. 
    \item Specifically, in this paper, we highlighted three key data integrity attacks against short term load forecasting FL models. We design the data integrity threat models and their counter measures.
    
    \item We further extend the proposed framework towards a privacy-preserving SIGNSGD-based FL approach whereby the clients locally perturb their trained parameters by adding noise prior to uploading to the server for aggregation to prevent parameter information leakage and ensure privacy preservation more effectively.
    
    \item We conduct comprehensive case studies and extensive empirical evaluations to verify the effectiveness of our proposed scheme using a real Australian energy consumption dataset obtained from Ausgrid Network.
\end{enumerate}

\noindent Table \ref{symbol} briefly introduces some commonly encountered symbols in our paper. The rest of this paper is structured as follows. Section \ref{sect:prelim} provides some background information in relation to our conceptual framework. Section \ref{sect:probdef} covers the problem definition section where we discuss some popular adversarial Byzantine threat models on FL. In Section \ref{propmethod}, we describe our proposed FL architecture followed by Section \ref{Results} which focusses on the evaluation and comparison of our proposed framework under several scenarios. Finally, Section \ref{Conclusion} concludes this manuscript and provides some potential future directions for research.

\begin{table}[!h]
\centering
\caption{Commonly used symbols \label{symbol}}
\begin{tabular}{|l|l|}

\hline
\textbf{Symbols} & \textbf{Definitions}       \\ \hline
$\eta$           & Learning rate              \\ \hline
$k$              & Client k                   \\ \hline
$N$              & Total number of clients    \\ \hline
$E_{k}$          & Paillier encryption scheme \\ \hline
$||.||$          & Encrypted parameter        \\ \hline
$T_{cl}$         & Communication round        \\ \hline
$D_k$            & Local dataset              \\ \hline
$\zeta_{k}$      & Gaussian noise             \\ \hline
$sign(.)$        & Sign vector                \\ \hline
\end{tabular}
\end{table}

\section{Related Works}

In what follows, we summarize the current state-of-the-art research on FL and Byzantine threats into two main categories as in the following:

\subsection{Federated Learning in load forecasting applications}

FL is a novel paradigm that enables collaborative training of machine learning models without requiring the transmission of data samples to a centralized server \cite{husnoo2022fedrep}. Since its inception, FL has been applied to several smart grid applications where privacy is paramount. Taïk and Cherkaoui \cite{9148937} first leveraged the application of FL in the load forecasting domain by training a LSTM model on a real-world Texas load consumption dataset and achieved sufficient forecasting performance. Another work by Venkataramanan et al. \cite{9729772} designed a FL-based framework for distributed energy resources forecasting and claimed high forecasting performance based on validations using GridLAB-D simulations and Pecan Street dataset. Similarly, the work in \cite{Yang_Wang_Zhao_Wu_2023} combined federated k-means clustering with variational mode decomposition and SecureBoost for short-term load forecasting. They claimed to achieve the lowest MAPEs of all existing algorithms for one-step ahead forecasting for both the US and Australian dataset used. Furthermore, the authors in \cite{9888131} proposed a federated hierarchical clustering solution to short-term load forecasting and claimed effectiveness and computational savings after validation on the Low Carbon London dataset. Similarly, a number of other related works \cite{Zhang_Zhu_Bai_2022, 9770488, Xu_Chen_Li_2022} have leveraged FL in the domain of load forecasting.

\subsection{Byzantine Threats}

Typically, Byzantine threats on FL scenarios consist of updating arbitrary model parameters from the clients to the server in the aim of impacting the convergence of the model \cite{BarossoFedThreatSurvey}. More specifically, Byzantine attacks are typically untargeted threats during which adversarial clients either train their local models on corrupted datasets or fabricate random model updates. Inherently, Byzantine threats are usually less stealthy and can be detected through close analysis of the global model performance \cite{9220780}. To address Byzantine resiliency in FL, a number of works have been proposed in recent literature. Throughout this section, we briefly summarize the main studies undertaken in regard to Byzantine resiliency in FL.

A common approach to Byzantine resiliency in FL is to employ  aggregation operators which are based on statistically robust estimators. For instance, the authors in \cite{FLTrust, 10.1145/3154503, 9029245} leveraged the use of Byzantine-robust aggregation rules by comparing the local updates of clients and filtering out statistical outliers prior to global model updates. Furthermore, Blanchard et al. \cite{10.5555/3294771.3294783} proposed a computationally expensive \textit{Krum} algorithm which performs gradient update selection and has the least sum of distances from the nearest gradient updates during each iteration. In addition, \cite{pmlr} introduced \textit{Bulyan} as an extension of Krum to recursively find subset of nodes using Krum and eventually perform an element-wise pruned mean on the updates to exclude the high magnitude values. The authors in \cite{distChen} propose a novel gradient correction strategy to solve the issue of non-convergence due disparity between the expected median and mean over the local gradients in heterogeneous settings by proposing a controlled noise perturbation scheme. Moreover, Li et al. \cite{RSALi} proposed a subgradient-based method termed as Byzantine-Robust Stochastic Aggregation (RSA) which does not rely upon the i.i.d. assumptions of data in client nodes. Similarly, a handful of other Byzantine-robust aggregation operators \cite{pmlr_v80_yin18a, 9153949, PillutlaAggregation, GonzalezByzantine, ShuhaoResidual} have been proposed in existing literature to mitigate the vulnerability of FL to Byzantine attacks. Another interesting study in \cite{9669031} utilized a mixed-strategy game-theoretic approach between the server and the clients whereby each client can either update good or corrupted model parameters while the server can either choose to accept or discard them. By employing the Nash Equilibrium property, the clients' updates were selected based on their probability of providing the correct updates.

In addition to the design of Byzantine-robust operators, several other defence strategies have been employed through anomaly detection \cite{ShiqiDefending, 9054676, 8975792}, pre-processing methods \cite{https://doi.org/10.48550/arxiv.2004.04986}, etc. It is worth noting that Byzantine robust statistical operators are designed to handle malicious or faulty data in federated systems. They can provide reliable estimates of statistical parameters even in the presence of attacks. On the other hand, Anomaly detectors can be used with these operators to identify malicious data points and improve the accuracy of data analysis in critical scenarios.However, while much work has been carried out to mitigate the threats of FL, little to no work has been carried out on secure, privacy-preserving and fault-tolerant FL frameworks for residential short-term electrical load forecasting to the best of our knowledge.  

\section{Preliminary}
\label{sect:prelim}
Throughout this section, we will discuss some preliminary and related background knowledge on FL and Differential Privacy (DP). Furthermore, within this section, we shall discuss a conventional FL set-up for short term load forecasting which will be used as a baseline during the evaluation of our proposed scheme.

\subsection{Federated Learning}
For the past couple of decades, Artificial Intelligence (AI) has transformed every walk of life and proven its benefits within several fields. However, one of the biggest real-world challenge faced by AI is the design of high-performing models due to natural data fragmentation coupled with security and privacy enforcement. Therefore, McMahan et al. \cite{Fedpap} introduced a fundamentally novel distributed learning concept which provides an alternative approach to leave the training data on the edge during learning. Specifically, the authors developed a collaborative decentralized on-device machine learning model training that does not require physical migration of raw data to a centralized server as compared to previous model training approaches and termed it as \textit{FL}. A brief example of FL is as shown in Fig. \ref{fig:fedillus} below:
\begin{figure}[!h]
    \centering
    \includegraphics[width=7cm]{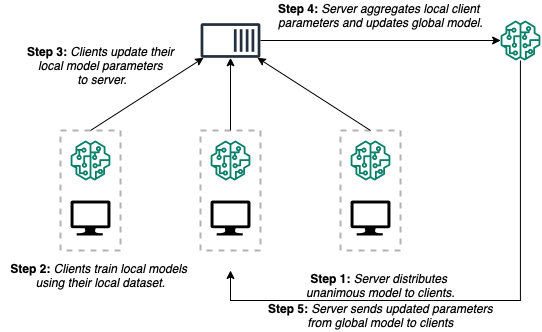}
    \caption{An illustration of the steps involved in FL.}
    \label{fig:fedillus}
\end{figure}

Suppose we have $N$ clients and each client $C_i$ holds a local training dataset $D_i$ where $i \in {1,2,...,N}$. An active $C_i$, participating in the local training, aims to collaboratively learn the weights $w_i$ of the shared global model such that a certain empirical loss $L_i$ is minimized. Therefore, we can formulate the optimization problem solved by multiple data owners as $w^* = \underset{w_i}{\mathrm{arg\ min}} \displaystyle \sum_{i=1}^N L_{i} (w_{i})$. Specifically, each communication round proceeds as shown in Fig. \ref{fig:fedillus} through the following steps: (1) The central server sends a unanimous global model $w$ to the active FL clients. (2) Each client trains the local model by using its own local dataset $D_i$ in order to solve the optimization problem $\underset{w_i}{\mathrm{min}}\ L_i(D_i, w_i)$. (3) Each client updates its local model parameters to the central server. (4) The server computes the global model update by aggregating the parameters received from the local models such that. (5) Lastly, the server sends back the updated parameters to the local models. This iterative process is continued until convergence of the global model. 

Furthermore, there are two baseline approaches to train models in a FL set-up namely Federated Averaging (FedAvg) and Federated Stochastic Gradient Descent (FedSGD). Generally, FedSGD \cite{pmlr-v119-malinovskiy20a} averages the locally computed gradient at every step of the learning phase while FedAvg \cite{9488877} averages local model updates when all the clients have completed training their models. However, as mentioned before, regardless of the approach used, FL is prone to several privacy and security threats, which have been discussed as following.

\subsection{Differential Privacy}
\label{sect:diffpriv}
Due to the several drawbacks of data anonymization techniques such as loss of data utility, risks of re-identification, etc., Differential Privacy (DP) emerged as a formal framework that enables the quantification of the preservation of individual privacy within a statistical database during the release of useful aggregate information \cite{9594795}. Therefore, we formally define some related concepts in relation to DP as in the following:

\noindent \textbf{Definition 1}: A randomized algorithmic mechanism $M: X \longrightarrow R$ with domain $X$ and range $R$ satisfies ($\epsilon$, $\delta$)-differential privacy if for all measurable sets $S \subseteq R$ and if for any two adjacent inputs $D$, $D' \in X$, the following holds: $Pr[M(D) \in S] \leq exp(\epsilon) \times Pr[M(D') \in S] + \delta$. Here $Pr$ denotes probability \cite{9594795}. Note that the parameters $\epsilon$ and $\delta$ are assumed to be positive real numbers. This is necessary to ensure that the privacy guarantee of the mechanism is meaningful and non-trivial \cite{9745062}.

\noindent \textbf{Definition 2}: The privacy loss $L$ of a randomized algorithmic mechanism $M: X \longrightarrow R$ for any result $v \in R$ and for any two data samples $D$, $D' \in X$ is expressed as: $L(v, D, D') = log \dfrac{Pr[M(D) = v]}{Pr[M(D') = v]}$. Privacy loss ensures that mechanisms are designed to protect the privacy of individuals' data while still providing useful and accurate results.\cite{9594795}

One of the most popular noise addition mechanisms for DP is the Gaussian Mechanism. A given noise distribution $n \sim N(0, \sigma^2)$ preserves ($\epsilon$,$\delta$)-DP where $N$ is a Gaussian distribution with 0 mean and variance $\sigma^2$, such that the noise scale is $\sigma \geq c\Delta s/\epsilon$ and the constant $c \geq \sqrt{2ln(1.25/\delta)}$ for $\epsilon \in (0,1)$ where $\Delta s$ is the sensitivity of the real-valued function. However, it is important to note that choosing the right amount of noise is a significant challenge that still lingers within research.

\subsection{Federated Load Forecasting with FedSGD (Benchmark)}

During FedSGD, a distributed stochastic gradient descent algorithm is applied within a federated environment to jointly train the global model. As shown in Algorithm \ref{FedSGDalgo}, our benchmark training algorithm uses FedSGD to update the parameters of our machine learning model. FedSGD is a stochastic algorithm because the local gradients computed on each client device are based on a random sample of the client's data, which is represented by the random subset of local dataset $D_k$ used to compute $g_k$ on each client. Moreover, the communication between the clients and the server is also subject to stochastic noise and delays. These sources of randomness and noise in the FedSGD algorithm can help to prevent the model from becoming too specialized to the training data and can improve its ability to make accurate predictions on new data. Therefore, in each communication round $T_{cl}$, we compute the stochastic gradient $g_k$ on each client using a random subset of its local data. We then send these local gradients to the Control Centre, where they are aggregated using FedSGD to obtain the global gradient $g$. Finally, the updated gradients are pushed back to the local models, and the process is repeated for the next round of training.

\begin{algorithm}
\textbf{Input}: learning rate $\eta$, each client $k$, local data $D_{k}$.

Control centre initializes and distributes unanimous model $m_0$ and encrypted parameter initialization $||\hat{m}_0||$ to all clients $N$.

\For{each communication round $T_{cl} = 1,2,..., t$}{

\For{each client $k \in N$}{

Compute stochastic gradient $g_k$ by training model on a random subset of local dataset $D_k$.

Send $g_k$ to Control Centre.

\textbf{end}
}
Control Centre aggregates the local gradient updates $g$ using FedSGD.

Control centre pushes updated gradients back to the local models.

\textbf{end}
}

\caption{Short-term Load Forecasting with FedSGD.}
\label{FedSGDalgo}
\end{algorithm}

%%%%%%
\subsection{Federated Load Forecasting with FedAvg (Benchmark)}

During FedAvg, a distributed averaging algorithm is applied within a federated environment to jointly train the global model. Our benchmark training algorithm uses FedAvg to update the parameters of the machine learning model. Unlike FedSGD, FedAvg computes the average of local model weights instead of gradients. As shown in Algorithm \ref{FedAvgalgo}, in each communication round $T_{cl}$, each client device trains the model on a random subset of its local data $D_k$ and computes its local weights $w_k$. These local weights are then sent to the Control Centre, where they are aggregated using FedAvg to obtain the global weights $w$. The updated global weights are then sent back to the local models for the next round of training. The averaging of the weights from multiple clients helps to reduce overfitting and improve the generalizability of the model. Furthermore, FedAvg is robust to client failures and non-i.i.d. data distribution among clients, which makes it suitable for FL in real-world scenarios load forecasting scenarios.

\begin{algorithm}
\textbf{Input}: learning rate $\eta$, each client $k$, local data $D_{k}$.

Control centre initializes and distributes unanimous model $m_0$ to all clients $N$.

\For{each communication round $T_{cl} = 1,2,..., t$}{

\For{each client $k \in N$}{

Compute gradient $g_k$ by training model on local dataset $D_k$.

Send $g_k$ to Control Centre.

\textbf{end}
}

Control Centre aggregates the local gradient updates $g$ using FedAvg:
$g = \frac{1}{|N|} \sum_{k \in N} w_k g_k$

where $w_k$ represents the weight of client $k$ which can be determined based on factors such as the number of samples in their local dataset.

Control centre pushes updated gradients back to the local models.

\For{each client $k \in N$}{

Update local model using aggregated gradient $g$:
$m_k = m_{k-1} - \eta g$

\textbf{end}
}
}
\caption{Short-term Load Forecasting with FedAvg.}
\label{FedAvgalgo}
\end{algorithm}

%%%%%%%%%%%%%%%%%%%%%%%%%%%%%%%%%%%

\section{Problem Definition \& Adversarial Models}
\label{sect:probdef}
FL enables promising privacy-preserving data analytics for smart grids by pushing model training to devices, thus requiring no direct data sharing \cite{9084352}. Nonetheless, recent literature has revealed its failure to sufficiently guarantee privacy preservation due to update leakage \cite{bhowmick2019protection}, deep leakage\cite{geng2022general}, Byzantine attacks \cite{247652}, etc. Throughout this paper, we aim to address Byzantine threats in relation to FL for electrical load forecasting. Before we present our proposed defense strategy, in this section, we consider three types of Byzantine threat models on federated load forecasting as in the following:
\begin{enumerate}[leftmargin=*]
    \item \textbf{Threat Model 1} \textit{(Local Data Poisoning)}: In this scenario, we assume that there is a subset of clients in the FL framework that are malicious or controlled by a malicious attacker. The malicious clients may have been introduced to the FL system through the addition of adversarially-controlled smart metering devices. The goal of the attacker is to manipulate the learned parameters of the global model in such a way that the model produces high indiscriminate errors. This implies that the attack objective is to maximize the sum of misclassifications on the poisoned samples: $Attack(D_{k} \cup D_{k}', m_{t}^k) = \max_{m} \sum_{i=1}^{n} [f(x_i'; m_{t}^k) \neq t_i']$. Here, $D_{k}'$ denotes the poisoned dataset of client $k$, $m_{t}^k$ represents the updated model after training on the poisoned samples, $x_i$ is the original sample, and $t_i'$ is the true label for the poisoned sample $x_i'$. Note that $f(x_{i}'; m_{t}^k)$ is the prediction of the model $m_{t}^k$ on the poisoned sample $x_{i}'$. The malicious clients can alter their local training data $D_k$ in a stealthy manner, but they cannot access or manipulate the data or training process of other clients or the global model. In this threat model, we assume that the attacker has complete knowledge of the training and validation data, as well as the training algorithm and the model architecture. Let $D_k = {(x_i, t_i)|i = 1,...,n}$ denote the pristine local training dataset with $n$ samples, where $x_i$ is the time instance and $t_i$ is the corresponding electrical load consumed. Each malicious client $k$ modifies their dataset $D_k$ by inserting a trigger $v$ into a random subset of training samples $x_i$, resulting in a poisoned dataset $D_{k}'$. Specifically, the poisoned sample $x_{i}'$ is obtained by adding the trigger $v$ to the original sample $x_i$, i.e., $x_{i}' = x_i + v, t_i$. The trigger $v$ can be a perturbation that is carefully designed to cause a specific outcome on the learned model, such as inducing a bias towards a certain class label. The attacker's ultimate goal is to degrade the accuracy of the global model on the test data, and potentially cause harm or disruption to the FL system. This threat model poses a significant challenge for FL systems, as it is difficult to detect and mitigate the effects of poisoned data on the global model.

    \item \textbf{Threat Model 2} \textit{(Model Leakage \& Poisoning)}: We consider the scenario where a subset of the clients participating in the FL framework are controlled by a malicious attacker who aims to poison the global model $M$ learned by the other clients. The attacker may be an insider who has legitimate access to the FL system or an outsider who manages to infiltrate it. The attacker's goal is to manipulate the learned parameters of the global model $M$ such that it performs poorly on specific tasks or adversarial examples, while still appearing to perform well on other tasks or data samples. This is achieved by injecting malicious updates into the model during the aggregation process, so that the updated model $m_t^k$ received by the aggregator contains the attacker's hidden agenda. Let $m_t^k$ represent the model update submitted by client $k$ at iteration $t$, and let $M_{t-1}$ denote the global model at the previous iteration. The attacker's goal is to craft a malicious model update $\Delta m_t^k$ that maximizes their attack objective, while still being able to fool the FL framework into accepting it as a legitimate update. Specifically, the attacker's objective can be formulated as $Attack(M_{t-1}, \Delta m_t^k) = \max L(M_t(M_{t-1}+\Delta m_t^k), D_{test})$, where $L$ is a loss function measuring the performance of the model $M_t$ on a test dataset $D_{test}$, and $M_t$ is the model learned by aggregating the updates ${m_t^k}_{k=1}^K$, including the attacker's malicious update $\Delta m_t^k$. The attacker may use various techniques to craft the malicious model update, such as gradient-based attacks, backdoor attacks, or model inversion attacks. The attacker may also try to evade detection by carefully choosing the magnitude and direction of the malicious update, or by injecting it into the updates of multiple clients in a coordinated manner. The attacker may also try to exploit any weaknesses in the FL framework, such as insecure communication channels, weak authentication, or untrusted aggregators.
    
    \item \textbf{Threat Model 3} \textit{(Colluding attack)}: In this scenario, a group of clients collude to perform a coordinated attack on the FL system. The colluding clients may be legitimate clients or may be controlled by a malicious attacker. The colluding clients work together to manipulate both their local training data and the model parameters to degrade the performance of the global model. The goal of the colluding attackers is to compromise the privacy and security of the system, extract sensitive information, or undermine the functionality of the FL system. The colluding attackers have access to their own local training datasets, $D_{k}$, and may work together to create a poisoned dataset, $D_{collusion}$, which is used to train the global model. The poisoned dataset may contain malicious samples that are designed to bias the model towards a particular outcome or to compromise the privacy of other clients. The colluding attackers may also work together to modify the model parameters in a coordinated way to achieve their goals. The colluding attackers will employ the two aforementioned techniques to perform their attack. The colluding attackers may also attempt to evade detection by the FL system by working in a stealthy and coordinated way. They may use encryption or other obfuscation techniques to conceal their actions from the FL system or may attempt to manipulate the FL system in a way that is difficult to detect.  Overall, the colluding attack threat model presents a significant challenge for FL systems, as it requires detecting and mitigating the actions of multiple attackers working in a coordinated way.
    
\end{enumerate}

%%%%%%%%%%%%%%%%%%%%%%%%%%%%%%%%

\section{Proposed Method}
\label{propmethod}

Within this section, we propose a new FL framework based on SIGNSGD, a gradient quantization mechanism, to circumvent the aforementioned Byzantine threats on FL for short-term load forecasting. The key idea lies in sharing just the sign of the gradients to preserve privacy. We present the our developed solution as in the following:

\subsection{System Model Overview}

\begin{figure}
    \centering
    \includegraphics[width=8cm]{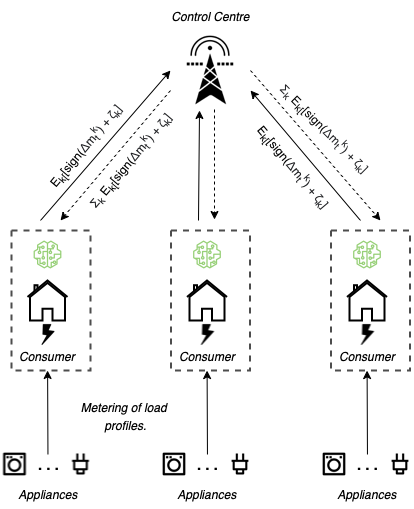}
    \caption{An illustration of proposed approach.}
    \label{fig:proposedapp}
\end{figure}

As previously discussed, the objective of this study is to design a robust and privacy-preserving FL framework for residential short-term load forecasting. As shown in Fig \ref{fig:proposedapp}, our proposed method  consists of three components as discussed.

\begin{enumerate}[leftmargin=*]
    \item \textit{Electrical Appliances}: Whenever someone within a household uses one of the electrical appliances, the load consumption is collected by the smart meter.
    \item \textit{Smart Meter}: Each customer has a smart meter that is connected a Home Area Network (HAN). Each smart meter collects energy load consumption profiles. The data collected is locally stored on the HAN of the consumer such that local models can be trained using their own dataset.
    \item \textit{Control Centre}: The control centre is responsible for broadcasting a learning model and default model parameters, aggregation of parameters after training and finally broadcasting the updated model parameters.
\end{enumerate}

\subsection{Gradient Quantization Mechanism}

The training of a LSTM-CNN model involves the optimization of an objective function $g(x)$ such that the loss is minimized. Typically, the search for the optimal network parameters to approximate $g(x)$ can be achieved through the use of FedSGD and FedAvg algorithms. As shown in Fig. \ref{fig:proposedapp}, large-scale training for federated load forecasting solutions requires frequent bi-directional communication of local gradient updates between the control center and HANs. However, the context of local model gradient can leak privacy through easy accessibility and exposes the FL framework to several Byzantine Threats, thus not conforming to the privacy-preserving guarantees of FL. Furthermore, the size of the local model updates also increases the communication bandwidth which limits scalability.

Therefore, we propose the use of a gradient quantization approach that approximates gradients to low-precision values. Specifically, we leverage  SIGNSGD whereby after local training, each client compresses its local gradient $g_k$ by utilizing a one-bit quantizer which simply takes the sign of $g_k$ such that $sign(g_k)$. Each client then sends its quantized gradients to the control center whereby aggregation occurs to obtain the one-bit compressed global gradient estimate as shown in Algorithm \ref{proposedalgo}. SIGNSGD is robust to byzantine attacks in FL setups in two ways. Firstly, it provides robustness to byzantine gradient updates by using only the direction of the gradient for updates, which ensures that the majority of correct worker nodes will determine the learning process. Secondly, it provides robustness to byzantine model poisoning by disregarding the magnitude of the gradient, which makes it impossible for byzantine worker nodes to influence the learning process by sending gradients with large magnitudes. Therefore, SIGNSGD helps ensure the accuracy and integrity of the learned model in the presence of adversarial behavior.

\subsection{Convergence Analysis}
In the following, we will present a formal analysis of the SIGNSGD approach through the use of refined assumptions derived from conventional SGD assumptions. 

\noindent\textbf{Assumption 1} \textit{(Lower Bound)}: Given a load forecasting objective function $f$, at any point $x$, $f(x) \geq f(x^)$, where $f(x^)$ represents the optimal objective value and $x^*$ represents the global minima of $f(x)$ over the federated dataset. This assumption is necessary to ensure the convergence to a stationary point \cite{jin2021stochasticsign}.

\noindent\textbf{Assumption 2} \textit{(Smoothness)}: Given a load forecasting objective function $f$, the gradient of $f$ (derivative of the function with respect to $x$) when evaluated on any coordinate $(x, y)$ can be represented as $g(x)$. Then, for $\forall x, y$ and for some non-negative Lipschitz constant $L_{i}$, we require that $|f(y) - [f(x) + g(x)^T (y-x)]| \leq \frac{1}{2} \sum_{i}L_{i}(y_{i} - x_{i})^2$, where $T$ is the sensitivity threshold. This assumption is essential to guarantee that the loss $l$ of $f$ is smooth and convergence of gradient descent algorithms \cite{jin2021stochasticsign}.

\noindent\textbf{Assumption 3} \textit{(Variance Bound)}: In a federated load forecasting setup, each client computes a stochastic gradient estimate $\hat{g}i(x)$ independently, and sends it to the server for aggregation. We assume that each $\hat{g}i(x)$ is an independent, unbiased estimate of the true gradient $g(x)$ with bounded variance per coordinate $\mathbb{E}[\hat{g}i(x)] = g(x), \quad \mathbb{E}[(\hat{g}i(x){j} - g(x){j})^2] \leq \sigma{j}^2$ where $\sigma{j}^2$ is the uniform variance bound for coordinate $j$, $\mathbb{E}$ represents the average of a random variable over its probability distribution and $g(x)$ is the true gradient to be estimated. This assumption is necessary to grasp the fundamental properties of stochastic optimization algorithms \cite{jin2021stochasticsign}.

\noindent\textbf{Assumption 4} \textit{(Gradient Noise)}: At any given point $x$, each component of the stochastic gradient vector, $\hat{g}_i(x)$, must have a unimodal distribution that is also symmetric about the mean. This assumption ensures that the addition of extra noise for the purpose of differential privacy does not skew the distribution and decrease utility \cite{jin2021stochasticsign}.

Under these assumptions, we have the following result:

\noindent \textbf{Theorem 1} \textit{(Non-convex convergence rate of SIGNSGD for federated load forecasting)}: Run Algorithm \ref{proposedalgo} for $K$ iterations under Assumptions 1 to 3. Set the learning rate as $\delta_k = \dfrac{1}{\sqrt{||L||_1 K}}$ where $n_k = K$. Let $N$ be the cumulative number of stochastic gradient
calls up to step $K$, i.e. $N = O(K^2)$ where $O(K^2)$ indicates that the rate of growth of the the total number of iterations $N$ with respect to $K$ is at most quadratic. Then we have $\mathbb{E}[\dfrac{1}{
K} \displaystyle \sum_{k = 0}^{K-1}||g_k||_1 ]^2 \leq \dfrac{1}{\sqrt{N}}[\sqrt{||L||1 } (f_0 - f* \dfrac{1}{2}) + 2||\sigma||1]$, where $L$ is the Lipschitz constant of the objective function, $f*$ is the optimal objective function value, $f_0$ is the initial objective function value, $g_k$ is the gradient of the objective function at the $k$-th iteration, and $\mathbb{E}$ represents the average of a random variable over its probability distribution.
Theorem 1 provides a convergence guarantee for SIGNSGD in the federated load forecasting setting. It states that after $K$ iterations, the expected value of the average of the $L_1$ norms of the gradients is upper bounded by a function of the Lipschitz constant $L$, the initial and optimal objective function values $f_0$ and $f_*$ respectively, the variance of the stochastic gradients $\sigma$, and the total number of stochastic gradient calls $N$. The learning rate $\delta_k$ in Theorem 1 is set to a decreasing schedule that ensures convergence. The cumulative number of stochastic gradient calls $N$ is upper bounded by $O(K^2)$, which means that the convergence rate of SIGNSGD is at most quadratic with respect to the number of iterations. Overall, Theorem 1 provides a formal convergence analysis for SIGNSGD in the context of federated load forecasting, which is useful for understanding the performance of the algorithm and tuning its hyperparameters.

\subsection{Algorithm Design}
Within a conventional FL setting with $N$ clients, at round $t$, a selected client $k \in N$ performs local gradient descent iterations $T_{gd}$ using a common broadcasted local model $m_{t-1}$ on its local training sample $D_{k}$ such that a new updated model $m_{t}^k$ is obtained. Each client $k$ then sends its updated parameters $\Delta m_{t}^k = m_{t}^k - m_{t-1}^k$ to the central orchestrator which in turn aggregates model updates from all $N$ clients $\forall k \in N$ such that $m_{t} = m_{t-1} + \sum_{k \in N} \dfrac{|D_{k}|}{\sum_{j} |D_{j}|} \Delta m_{t}^k$. The model training continues until convergence and is subsequently terminated after a set number of rounds $T_{cl}$. 

\begin{algorithm}
\textbf{Input}: learning rate $\eta$, each client $k$ local data $D_{k}$.

Control centre distributes unanimous model $m_{0}$ and encrypted parameter initialization $||\hat{m_{0}}||$ to all clients $N$.

\For{each communication round, $T_{cl} =1,..., t$}{

\For{each client $k$}{

Compute the gradient $g_{k}= m_{t}^k - m_{t-1}^k$ by training on local dataset $D_k$.

Obtain sign vector $sign(\Delta m_{t}^k)$ from $g_k.$

Perturb $sign(\Delta m_{t}^k)$ with a random Gaussian noise $\zeta_{k}$ such that $\sum_{k \in N} sign(\Delta m_{t}^k) + \zeta_{k}$ satisfies differential privacy.

Encrypt $sign(\Delta m_{t}^k) + \zeta_{k}$ into $E_{k}[sign(\Delta m_{t}^k) + \zeta_{k}]$ and send to control centre.

\textbf{end}
}

Control Centre aggregates encrypted updates $\sum_{k} E_{N_{k}} (sign(\Delta m_{t}^k) + \zeta_{k})$.

Control Centre pushes $sign(g_N)$ to all clients, $N$. 

\textbf{end}
}
\caption{Proposed Framework.}
\label{proposedalgo}
\end{algorithm}

However, in the context of smart grids, conventional FL settings pose several privacy risks as earlier discussed. Therefore, we propose a novel privacy-preserving FL framework for electrical load forecasting that leverages the idea of gradient quantization mechanism. Following the inspiration from the wide adoption of advanced solvers in FL, the authors in \cite{jin2021stochasticsign} pioneered an intuitive and theoretically-sound method known as SIGNSGD which is a sign-based gradient quantization scheme for 1-bit compression and transmission of the sign of the gradient which in turn improves privacy and communication efficiency. Specifically, as shown in Algorithm \ref{proposedalgo} and Figure \ref{fig:proposedapp}, a selected client $k$ initially computes the gradient update $g_{k} = m_{t}^k - m_{t-1}^k$ from which it obtains the sign vector $sign(\Delta m_{t}^k) = sign(m_{t}^k - m_{t-1}^k)$ where $sign(\Delta m_{t}^k)$: $\mathbb{R}^n \longrightarrow {-1,1}^n$. A random Gaussian noise $\zeta_{k}$ is then added to perturb $sign(\Delta m_{t}^k)$ such that $\sum_{k \in N} sign(\Delta m_{t}^k) + \zeta_{k}$ satisfies differential privacy. Furthermore, to prevent an adversary from learning $sign(\Delta m_{t}^k) + \zeta_{k}$ accurately in circumstances where $N$ is large, each client $k$ updates the encrypted results $E_{k}[sign(\Delta m_{t}^k) + \zeta_{k}]$ to the central aggregator. Note that in our proposed framework, we make use of Paillier encryption scheme $E_{k}$ as it is 1) non-interactive, meaning that the encryption process does not require communication between the parties. This is important for FL setups, where the data is distributed across multiple devices or users, and communication between them can be slow or unreliable, and, 2) partially homomorphic which means that it allows the central aggregator to perform calculations on the encrypted updates without needing to decrypt them, which preserves the privacy of the data. The orchestrator in turn sums all the encrypted model updates from $N$ such that $\sum_{k} E_{N_{k}} (sign(\Delta m_{t}^k) + \zeta_{k})$. This aggregation follows the selection of the median of all $N$ clients  signs at every position of the update vector. The model training continues until convergence and is subsequently terminated after a set number of rounds $T_{cl}$. 

\subsection{Robustness Guarantees of SIGNSGD to Byzantine Attacks}

In what follows, we will present formal mathematical guarantees of the robustness of SIGNSGD to both Local Data Poisoning (Threat 1) and Model Leakage and Poisoning (Threat 2) Byzantine threats as in the following:

\noindent\textbf{Assumptions}: Let there be $k$ client nodes with i.i.d local datasets. The clients are honest-but-curious, meaning that they follow the FL protocol but may try to learn information about other nodes' data or influence the training process. We assume that at most $t$ clients may deviate from the protocol with byzantine behavior. The control centre uses a weighted average of signed gradient updates and broadcasts updates to the clients at the end of each iteration. We define the objective function $f(\theta)$ as the empirical risk of the model on the global dataset.

\subsubsection{Robustness to Local Data Poisoning Byzantine Attacks} Assume that there are $t$ clients that may send arbitrary updates to the control centre, which may contain poisoned data. We assume that the objective function is $L$-smooth, meaning its gradient is Lipschitz continuous with constant $L$. We further assume that the size of the poisoned data is bounded by a fraction of $\delta$ in each client's local dataset. We define the deviation of the model parameters from the optimal solution as $\delta_t = |\theta_t - \theta^*|$.

\noindent\textbf{Theorem 2}: If the fraction of poisoned data $\delta$ satisfies $\delta < \frac{1}{8L(t+1)}$, and the step size satisfies $\eta < \frac{1}{4L}$, then the deviation of the model parameters from the optimal solution caused by the byzantine local data poisoning is bounded as follows: $\delta_T \leq (1 + L\eta)^T . \left[\left(1 + \dfrac{2(t+1)\delta}{\sqrt{n}}\right)^{\sqrt{n}} + \dfrac{1}{\sqrt{n}}\right]$. Note that this bound depends on the number of byzantine worker nodes $t$, the Lipschitz constant $L$ of the objective function, the step size $\eta$, and the size of the poisoned data \cite{yang2019byzantine}.

\subsubsection{Robustness to Model Poisoning Byzantine Attacks} Assume that there are $t$ worker nodes that may be byzantine and try to poison the model by sending arbitrary gradients. We further assume that the model is $M$-bounded, meaning that its parameters are bounded in magnitude by $M$. We define the deviation of the model parameters from the optimal solution as $\delta_t = |\theta_t - \theta^*|$.

\noindent\textbf{Theorem 3}: If the number of byzantine worker nodes $t$ satisfies $t < \frac{n}{4}$, and the step size $\eta$ satisfies $\eta < \frac{1}{L}$, then the deviation of the model parameters from the optimal solution caused by the byzantine model poisoning is bounded as follows: $\delta_T \leq \left( 1 + \dfrac{2M^2 \eta t}{n}\right)^T . \: \delta_0 + \dfrac{2M^2 \eta}{L} . \left(1 + \dfrac{2M^2 \eta t}{n}\right)^T . \: \sqrt{\dfrac{8(t+1)log \, n}{n}}$. Note that this bound depends on the number of byzantine worker nodes $t$, the $M$-bound of the model, the step size $\eta$, and the size of the deviation of the model parameters from the optimal solution at time $t=0$ \cite{pmlr-v80-yin18a}.

\section{Simulation \& Results}
\label{Results}
In this section, we provide the results of the experimental evaluations of our proposed approach. We first introduce the dataset used and the settings shared by all experiments. Next, the performance of the proposed approach is presented and compared throughout different scenarios. Lastly, we discuss the overall results.

\subsection{Experimental Setup}

This research was conducted using \textit{Solar Home Electricity Data} from Eastern Australia's largest electricity distributor, Ausgrid. The dataset composes of half-hourly electricity consumption data of 300 de-identified customers which is measured using gross meters during the period starting 1\textsuperscript{st} July 2012 to 30\textsuperscript{th} June 2013. We initially filter the data based on General Consumption (GC) category. It is then converted to the suitable time-series format. It is then split into test (30\%) and train (70\%) subsets. 

Every experiment carried out have the following general configurations. There is a set number of clients (10 clients) each holds a local subset of the data and there is a server which helps to coordinate the FL scenario. Specifically, in the experiments, we used 210 customer data points for training our short-term load forecasting model. To ensure a fair distribution of the data across the client nodes in our FL setup, we randomized the data and assigned an equal amount to each of the 10 client nodes. We also introduced randomness to the data to account for the fact that different customers may have unique and stochastic load profiles. By training our model on this randomized data, we aimed to improve its ability to generalize to new and unseen load profiles, ultimately leading to better forecasting accuracy. The model performance is evaluated using three metrics: \textit{Mean Squared Error (MSE)}, \textit{Root Mean Squared Error (RMSE)} and lastly, \textit{Mean Absolute Percentage Error (MAPE)}.

\subsection{Comparison with Baseline (No Attack)}
\begin{figure}[!h]
    \centering
    \subfloat[\centering Convergence of Federated LSTM-CNN model. ]{{\includegraphics[width=4cm]{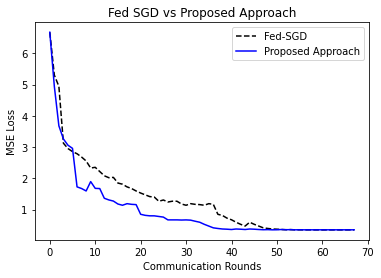} }}%
    \qquad
    \subfloat[\centering MAPE (\%) per client. ]{{\includegraphics[width=3.9cm]{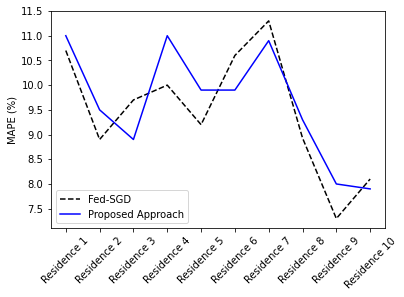} }}%
    \caption{Comparison between FedSGD and proposed approach.}%
    \label{fig:trainloss}
\end{figure}

Throughout this section, we present the experimental results to compare the performance of the proposed approach against the conventional FedSGD approach. As shown in Fig. \ref{fig:trainloss}(a), it can be seen that the FedSGD reaches convergence after the 47\textsuperscript{th} communication round while the proposed approach converges after the 40\textsuperscript{th} communication round.
\begin{table}[!h]
\caption{Evaluation of FedSGD with several models. \label{CompaGedSGD}}
\begin{tabular}{|l|l|l|l|l|l|}
\hline
\textbf{Metric}    & \textbf{RNN} & \textbf{GRU} & \textbf{LSTM} & \textbf{CNN} & \textbf{LSTM-CNN} \\ \hline
\textbf{MSE}       & 0.2657       & 0.1973       & 0.1634        & 0.2567       & 0.1583            \\ \hline
\textbf{RMSE}      & 0.5346       & 0.4042       & 0.3463        & 0.5243       & 0.3008            \\ \hline
\textbf{MAPE (\%)} & 16.4         & 10.9         & 11.0          & 12.8         & 9.7               \\ \hline
\end{tabular}
\end{table}

\begin{table}[!h]
\caption{Evaluation of proposed method with several models. \label{CompaSignSGDModel}}
\begin{tabular}{|l|l|l|l|l|l|}
\hline
\multicolumn{1}{|l|}{\textbf{Metric}}    & \multicolumn{1}{l|}{\textbf{RNN}} & \multicolumn{1}{l|}{\textbf{GRU}} & \multicolumn{1}{l|}{\textbf{LSTM}} & \multicolumn{1}{l|}{\textbf{CNN}} & \multicolumn{1}{l|}{\textbf{LSTM-CNN}} \\ \hline
\multicolumn{1}{|l|}{\textbf{MSE}}       & \multicolumn{1}{l|}{0.2662}       & \multicolumn{1}{l|}{0.1864}       & \multicolumn{1}{l|}{0.1803}        & \multicolumn{1}{l|}{0.2456}       & \multicolumn{1}{l|}{0.1437}            \\ \hline
\multicolumn{1}{|l|}{\textbf{RMSE}}      & \multicolumn{1}{l|}{0.5432}       & \multicolumn{1}{l|}{0.4127}       & \multicolumn{1}{l|}{0.3890}        & \multicolumn{1}{l|}{0.5329}       & \multicolumn{1}{l|}{0.3243}            \\ \hline
\multicolumn{1}{|l|}{\textbf{MAPE (\%)}} & \multicolumn{1}{l|}{15.9}         & \multicolumn{1}{l|}{11.1}         & \multicolumn{1}{l|}{10.8}          & \multicolumn{1}{l|}{13.6}         & \multicolumn{1}{l|}{9.7}                           \\ \hline      
\end{tabular}
\end{table}

\begin{table*}[!ht]
\centering
\caption{Evaluation of proposed FL framework against Threat Model 1 \& 2. \label{EvaluationFLThreat}}
\begin{tabular}{cc|cc|cc|}
\cline{3-6}
\multicolumn{2}{l|}{\textbf{}}                                                                                          & \multicolumn{2}{c|}{\textbf{FedSGD}}                                  & \multicolumn{2}{c|}{\textbf{Proposed Solution}}                                             \\ \hline
\multicolumn{1}{|c|}{\textbf{\begin{tabular}[c]{@{}c@{}}\% of Compromised\\ Clients\end{tabular}}} & \textbf{Metric}    & \multicolumn{1}{c|}{\textbf{Threat Model 1}} & \textbf{Threat Model 2}  & \multicolumn{1}{l|}{\textbf{Threat Model 1}} & \multicolumn{1}{l|}{\textbf{Threat Model 2}} \\ \hline
\multicolumn{1}{|c|}{\multirow{3}{*}{\textbf{10}}}                                                 & \textbf{MSE}       & \multicolumn{1}{c|}{0.2910}                  & 0.3134                  & \multicolumn{1}{c|}{0.1621}                  & 0.1532                                       \\ \cline{2-6} 
\multicolumn{1}{|c|}{}                                                                             & \textbf{RMSE}      & \multicolumn{1}{c|}{0.4732}                  & 0.5490                  & \multicolumn{1}{c|}{0.3251}                  & 0.3029                                       \\ \cline{2-6} 
\multicolumn{1}{|c|}{}                                                                             & \textbf{MAPE (\%)} & \multicolumn{1}{c|}{18.2}                    & 20.1                    & \multicolumn{1}{c|}{10.1}                    & 9.9                                          \\ \hline
\multicolumn{1}{|c|}{\multirow{3}{*}{\textbf{20}}}                                                 & \textbf{MSE}       & \multicolumn{1}{c|}{0.4180}                  & 0.4519                  & \multicolumn{1}{c|}{0.1835}                  & 0.1642                                       \\ \cline{2-6} 
\multicolumn{1}{|c|}{}                                                                             & \textbf{RMSE}      & \multicolumn{1}{c|}{0.7893}                  & 0.9201                  & \multicolumn{1}{c|}{0.3502}                  & 0.3129                                       \\ \cline{2-6} 
\multicolumn{1}{|c|}{}                                                                             & \textbf{MAPE (\%)} & \multicolumn{1}{c|}{25.7}                    & 27.1                    & \multicolumn{1}{c|}{12.2}                    & 10.8                                         \\ \hline
\multicolumn{1}{|c|}{\multirow{3}{*}{\textbf{30}}}                                                 & \textbf{MSE}       & \multicolumn{1}{c|}{0.7319}                  & 0.8192                  & \multicolumn{1}{c|}{0.2678}                  & 0.2134                                       \\ \cline{2-6} 
\multicolumn{1}{|c|}{}                                                                             & \textbf{RMSE}      & \multicolumn{1}{c|}{1.2398}                  & 1.4576                  & \multicolumn{1}{c|}{0.4249}                  & 0.3965                                       \\ \cline{2-6} 
\multicolumn{1}{|c|}{}                                                                             & \textbf{MAPE (\%)} & \multicolumn{1}{c|}{38.9}                    & 42.2                    & \multicolumn{1}{c|}{17.3}                    & 14.1                                         \\ \hline
\end{tabular}
\end{table*}

\begin{table}[]
\caption{Evaluation of proposed FL framework against Threat Model 3. \label{EvaluationFLThreat3}}
\begin{tabular}{|c|c|c|c|}
\hline
\multicolumn{1}{|l|}{\textbf{\% of Comp. Clients}} & \multicolumn{1}{l|}{{ \textbf{Metric}}} & \multicolumn{1}{l|}{{\textbf{FedSGD}}} & \multicolumn{1}{l|}{{\textbf{Proposed Solution}}} \\ \hline
                                                         & \textbf{MSE}                                                & 0.3103                                                       & 0.1732                                                                 \\ \cline{2-4} 
                                                         & \textbf{RMSE}                                               & 0.5321                                                       & 0.3324                                                                 \\ \cline{2-4} 
\multirow{-3}{*}{\textbf{20}}                            & \textbf{MAPE (\%)}                                          & 19.3                                                         & 11.2                                                                   \\ \hline
                                                         & \textbf{MSE}                                                & 0.5231                                                       & 0.2034                                                                 \\ \cline{2-4} 
                                                         & \textbf{RMSE}                                               & 0.8743                                                       & 0.3958                                                                 \\ \cline{2-4} 
\multirow{-3}{*}{\textbf{30}}                            & \textbf{MAPE (\%)}                                          & 34.0                                                         & 14.0                                                                   \\ \hline
                                                         & \textbf{MSE}                                                & 0.7793                                                       & 0.2901                                                                 \\ \cline{2-4} 
                                                         & \textbf{RMSE}                                               & 1.2343                                                       & 0.4302                                                                 \\ \cline{2-4} 
\multirow{-3}{*}{\textbf{40}}                            & \textbf{MAPE (\%)}                                          & 39.5                                                         & 16.4                                                                   \\ \hline
\end{tabular}
\end{table}

\begin{table}[]
\centering
\caption{Evaluation of proposed FL framework under different Privacy Budgets. \label{threat3}}
\begin{tabular}{|c|c|c|c|}
\hline
\textbf{$\epsilon$-Budget}     & \textbf{Metric}    & \textbf{FedSGD} & \multicolumn{1}{l|}{\textbf{Proposed Solution}} \\ \hline
\multirow{3}{*}{\textbf{0.01}} & \textbf{MSE}       & 0.1583           & 0.1437                                          \\ \cline{2-4} 
                               & \textbf{RMSE}      & 0.3              & 0.3243                                          \\ \cline{2-4} 
                               & \textbf{MAPE (\%)} & 9.7              & 9.7                                             \\ \hline
\multirow{3}{*}{\textbf{0.1}}  & \textbf{MSE}       & 0.4320           & 0.1645                                          \\ \cline{2-4} 
                               & \textbf{RMSE}      & 0.8173           & 0.3192                                          \\ \cline{2-4} 
                               & \textbf{MAPE (\%)} & 26.4             & 10.5                                                      \\ \hline
\end{tabular}
\end{table}

\begin{figure*}[!h]
    \centering
    \subfloat[\centering Impact of Threat Model 1. ]{{\includegraphics[width=5cm]{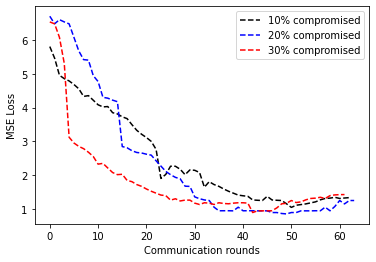} }}%
    \subfloat[\centering Impact of Threat Model 2. ]{{\includegraphics[width=5cm]{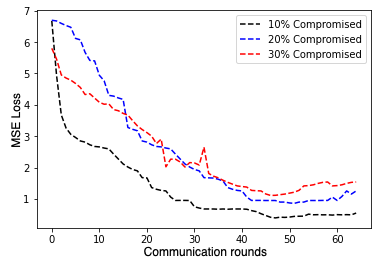} }}%
    \subfloat[\centering Impact of Threat Model 3.  ]{{\includegraphics[width=5cm]{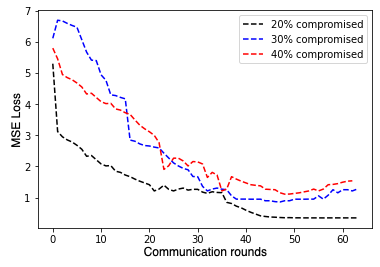} }}%
    \caption{Impact of Attacks on FedSGD.}%
    \label{fig:impactFedSGD}
\end{figure*}

\begin{figure*}[!h]
    \centering
    \subfloat[\centering Impact of Threat Model 1. ]{{\includegraphics[width=5cm]{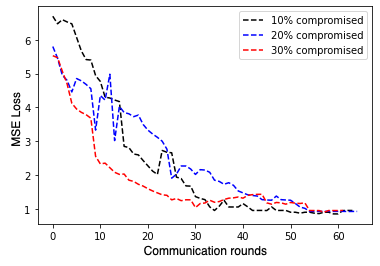} }}%
    \subfloat[\centering Impact of Threat Model 2. ]{{\includegraphics[width=5cm]{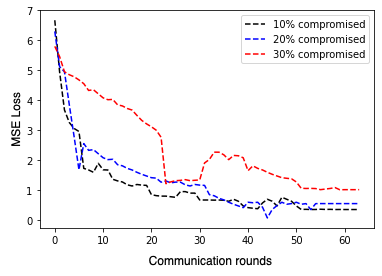} }}%
    \subfloat[\centering Impact of Threat Model 3.  ]{{\includegraphics[width=5cm]{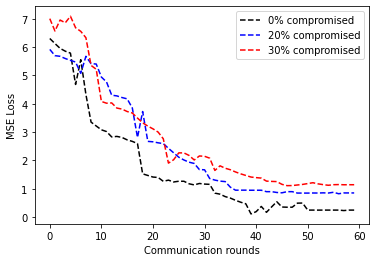} }}%
    \caption{Mitigating threat models using our proposed method.}% 
    \label{fig:impactSIGNSGD}
\end{figure*}

As our proposed solution converges faster that the traditional FedSGD one, we can conclude that the proposed approach provides a fast algorithmic convergence. Furthermore, we use the three aforementioned evaluation metrics to compare and contrast the performance of the proposed solution against FedSGD with several models as presented in Table \ref{CompaGedSGD} and Table \ref{CompaSignSGDModel}. The experimental results reveal that the the proposed framework reaches similar performance as compared to the FedSGD approach.  Similarly, in Fig. \ref{fig:trainloss}(b), the MAPE per active household within the FL set ups are contrasted which shows that our proposed approach reaches relatively similar performance as compared to the FedSGD. More specifically, after the comparison, we can deduce that our proposed framework reaches good generalization performance for short-term load forecasting within acceptable error ranges. Moreover, after comparing the proposed framework based on models as presented in Table \ref{CompaSignSGDModel}, it can be deduced that LSTM-CNN model shows the best overall forecasting performance with an average MAPE of 9.7\% in both the conventional FedSGD and the proposed FL framework. 

\subsection{Resilience of SIGNSGD to Byzantine Attacks}

In this section, we evaluate the robustness of our proposed federated load forecasting framework with SIGNSGD against the three adversarial threat models, namely local data poisoning, model leakage \& poisoning and, lastly, colluding threats, as discussed in Section \ref{sect:probdef}. We compare the performance of our proposed SIGNSGD approach against that of the benchmark FedSGD under all three scenarios. We use three metrices, namely Mean Squared Error (MSE), Root Mean Squared Error (RMSE) and Mean Absolute Percentage Error (MAPE) to evaluate the performance of our framework. Furthermore, we evaluate the performance of our proposed approach under three scenarios based on the number of compromised client nodes. Each experiment is carried out for five times and the average of the results is reported. Table \ref{EvaluationFLThreat} summarizes the comparison of our proposed solution and FedSGD under local data poisoning (threat model 1) and model poisoning (threat model 2). As expected, there is a significant decrease in the performance of FedSGD with increasing number of compromised clients under either threat model 1 and threat model 2. However, as compared with FedSGD, the performance of SIGNSGD in load forecasting scenario when under attack by threat model 1 or threat model 2 stays relatively stable with insignificant increases in MSE, RMSE and MAPE even with increasing number of adversarial client nodes. Similarly, as depicted in Table \ref{EvaluationFLThreat3}, our proposed framework with SIGNSGD differs from the benchmark FedSGD setup under colluding attacks (threat model 3) such that the change MSE, RMSE and MAPE are statistically insignificant, indicating that it is able to maintain good performance even in the presence of Byzantine attacks. 

Furthermore, we evaluate impact of the attacks on FedSGD-based load forecasting scenarios. Figure \ref{fig:impactFedSGD} shows the convergence of FedSGD under three different types of Byzantine attacks and different level of attack intensities. We can observe that as the number of Byzantine workers increases, the convergence of FedSGD deteriorates rapidly. This indicates that FedSGD is highly susceptible to Byzantine attacks, which can significantly degrade its performance in a real-world setting. The divergence in the convergence curves of FedSGD under Byzantine attacks highlights the importance of developing robust federated learning algorithms. Therefore, from Figure \ref{fig:impactSIGNSGD}, we  observe that the train loss values of our proposed approach using SIGNSGD remain fairly stable throughout the training process even with the presence of Byzantine nodes, indicating that SIGNSGD is robust to Byzantine attacks. Unlike FedSGD, the loss values of SIGNSGD do not experience divergence even when the number of adversarial client nodes increases. This validates that SIGNSGD is not only resilient to Byzantine attacks but also maintains its convergence properties even under the presence of malicious nodes.

\subsection{Computational Efficiency of Proposed Method}

\begin{figure}
    \centering
    \includegraphics[width=8cm]{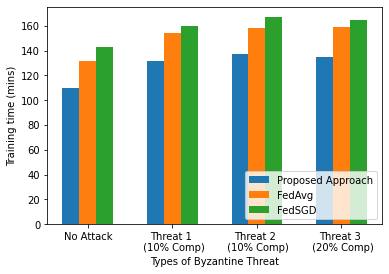}
    \caption{Comparison of Computational Efficiency.}
    \label{fig:computime}
\end{figure}

Subsequently, we evaluate the computational efficiency achieved by our proposed SIGNSGD method in comparison against the baseline. Specifically, we contrast the training time of our proposed FL framework against FedAvg and FedSGD, all of which are trained on similar data with the LSTM-CNN model. Similar model is used in aim of easing comparison such that the training time can be used as an estimate of the computational efficiency of our proposed approach. As depicted in Figure \ref{fig:computime}, we note that overall, our proposed approach has a lower computation time as compared to FedAvg and FedSGD. During attacks, there is an increase in the computation time in all cases. Indeed, it can be highlighted that the training time for our proposed approach is roughly around 20\% less than FedAvg and 30\% lower than that of FedSGD. This is due to the decrease in the byte size of the gradients shared to the control centre during training. 

It should be noted that the computation of updated model parameters does not need to be done online during the load forecasting execution. The training of  the LSTM-CNN model can happen off-line based on the training inputs that were collected during the forecasting execution, but at another time. In other words, there will be two separate algorithms/codes: one for training the LSTM-CNN parameters, which is carried out at any suitable time interval off-line, and another for executing the short-term load forecasting, which uses the trained LSTM-CNN online during the operation of the system. Therefore, the computation time related to the model training is not critical in the sense that it does not have to be done within a short time period.

\subsection{Results Discussion}

With increasing concerns and regulation enforcement in regards to security and privacy within the smart grid paradigm, it is crucial to develop privacy-preserving and robust short term load forecasting solutions. FL, whilst still being in its infant stage, requires further improvements under different circumstances. Therefore, throughout this study, we investigate Byzantine attacks in relation to federated short term load forecasting. Furthermore, we propose and design a robust defense solution to mitigate those threats. The proposed federated load forecasting framework with SIGNSGD was evaluated against three adversarial threat models: local data poisoning, model leakage \& poisoning, and colluding threats. The performance of SIGNSGD was compared to that of benchmark FedSGD, and the evaluation was carried out under three scenarios based on the number of compromised client nodes. The results in Tables \ref{EvaluationFLThreat} \&
\ref{EvaluationFLThreat3} as well as Figures \ref{fig:impactFedSGD} \&
\ref{fig:impactSIGNSGD} showed that SIGNSGD was relatively stable and maintained good performance even in the presence of Byzantine attacks, while the performance of FedSGD deteriorated rapidly with increasing number of adversarial client nodes. Furthermore, he computational efficiency achieved by SIGNSGD was also evaluated, and the training time was found to be roughly 20\% less than FedAvg and 30\% lower than that of FedSGD, due to the decrease in the byte size of the gradients shared during training. These findings indicate that SIGNSGD is a robust and computationally efficient federated learning algorithm that could be useful for applications where data privacy is important and the risk of malicious attacks is high such as in the energy and smart grid critical infrastructure domain.

\section{Conclusion}
\label{Conclusion}

The rapid adoption of FL within the smart grid ecosystem has spiked the interest of researchers to address its security and privacy issues. Byzantine attack mitigation plays a crucial role in securing and enhancing the robustness of FL for short-term load forecasting.  Therefore, throughout this manuscript, we propose a state-of-the-art FL-based approach that leverages the notions of gradient quantization and differential privacy to overcome this challenge. Furthermore, we empirically demonstrate that our proposed solution effectively mitigate popular Byzantine threats and provides relatively similar performance as compared to standard FL setups.  Finally, the next steps in this research are to: (1) design and evaluate our proposed FL framework against stronger Byzantine attacks, and, (2) take into consideration the existence of distributed energy resources to improve the grid model.  

\bibliographystyle{IEEEtran}
\bibliography{refs.bib}


% \end{IEEEbiography}


% % insert where needed to balance the two columns on the last page with
% %\newpage



%\vfill


\end{document}